\documentclass{ws-ijmpe}

\begin{document}

\markboth
{M. G\'o\'zd\'z and W. A. Kami\'nski} 
{Fermion-Boson Loops with Bilinear $R$-Parity Violation...}

%
%

\title{FERMION-BOSON LOOPS WITH BILINEAR $R$-PARITY \\ VIOLATION
  LEADING TO MAJORANA NEUTRINO MASS \\ AND MAGNETIC MOMENTS}

\author{MAREK G\'O\'ZD\'Z and WIES{\L}AW A. KAMI\'NSKI}

\address{
Department of Informatics, Maria Curie-Sk{\l}odowska University, \\
pl. Marii Curie--Sk{\l}odowskiej 5, 20-031 Lublin, Poland \\
mgozdz@kft.umcs.lublin.pl, kaminski@neuron.umcs.lublin.pl
}

\maketitle

\begin{history}                 %
\end{history}                  	%

\begin{abstract}
  We present analytic expressions corresponding to a~set of one loop
  Feynman diagrams, built within $R$-parity violating (RpV) minimal
  supersymmetric standard model (MSSM). Diagrams involve both bilinear
  and trilinear RpV couplings and represent Majorana neutrino masses and
  magnetic moments.
\end{abstract}

\section{Supersymmetric model with $R$-parity violation}

Building the minimal supersymmetric (SUSY) version of the Standard Model
(MSSM)\cite{mgozdz:mssm} one usually assumes the conservation of the
$R$-parity, defined as $R=(-1)^{3B+L+2S}$. In theories preserving the
$R$-parity the lepton and baryon numbers are conserved, and SUSY
particles are not allowed to decay to non-SUSY ones. It implies that the
lightest SUSY particle must remain stable, giving a~good natural dark
matter candidate. The $R$-parity conservation can be achieved by
neglecting certain theoretically allowed terms in the superpotential. By
retaining these terms one finishes with an~$R$-parity violating (RpV)
model,\cite{mgozdz:1RpV,mgozdz:2RpV,mgozdz:3RpV} with richer
phenomenology and many exotic interactions. The RpV models provide
mechanisms of generating Majorana neutrino masses and magnetic moments,
describe neutrino decay, SUSY particles decays, exotic nuclear processes
like the neutrinoless double beta decay, and many more. Being
theoretically allowed, RpV SUSY theories are interesting tools for
studying the physics beyond the Standard Model.

The $R$-parity conserving part of the superpotential of MSSM is usually
written as
\begin{eqnarray}
  W^{MSSM} &=& \epsilon_{ab} [(\mathbf{Y}_E)_{ij} L_i^a H_1^b \bar E_j
    + (\mathbf{Y}_D)_{ij} Q_{ix}^{a} H_1^b \bar D_{j}^{x} \nonumber \\
    &+& (\mathbf{Y}_U)_{ij} Q_{i x}^{a} H_2^b \bar U_{j}^{x} 
    + \mu H_1^a H_2^b], 
\end{eqnarray}
while its RpV part reads
\begin{eqnarray}
  W^{RpV} &=& \epsilon_{ab}\left[
    \frac{1}{2} \lambda_{ijk} L_i^a L_j^b \bar E_k
    + \lambda'_{ijk} L_i^a Q_{jx}^{b} \bar D_{k}^{x} \right] \nonumber \\
  &+& \frac{1}{2}\epsilon_{xyz} \lambda''_{ijk}\bar U_i^x\bar
  D_j^y \bar D_k^z + \epsilon_{ab}\kappa^i L_i^a H_2^b.
\end{eqnarray}
The {\bf Y}'s are 3$\times$3 Yukawa matrices. $L$ and $Q$ are the
$SU(2)$ left-handed doublets while $\bar E$, $\bar U$ and $\bar D$
denote the right-handed lepton, up-quark and down-quark $SU(2)$
singlets, respectively. $H_1$ and $H_2$ mean two Higgs doublets. We have
introduced color indices $x,y,z = 1,2,3$, generation indices
$i,j,k=1,2,3$ and the SU(2) spinor indices $a,b = 1,2$. One of the
major problems in these models is the assurance of proton stability,
therefore the baryon number violating coupling constant $\lambda''$ is
usually set to zero. It is also customary to neglect the bilinear
($\kappa^i$) term and concentrate on the phenomenology of the remaining
trilinear terms.

In the present paper we introduce processes which involve both bi- and
trilinear coupling constants. We focus on one loop Feynman diagrams
which potentially lead to Majorana neutrino masses and magnetic moments.
Such diagrams have been proposed by J.~D.~Vergados\cite{mgozdz:JDV} and,
to our best knowledge, this is the first attempt to treat them in an
exact analytical way. In the next section we present the full set of new
Feynman diagrams within the RpV MSSM model, together with the
corresponding analytical functions. We do not have any numerical results
yet. The work on a~stable renormalization group based code, which would
include the RpV effects, is in progress.

One of the characteristic features of the MSSM is that the lepton
doublet superfield $L$ has the same quantum numbers as the $H_1$ Higgs
doublet superfield. Thus in the RpV models one can combine them into one
superfield ${\cal L}_\alpha^a=\{H_1^a,L_{1,2,3}^a\},\ \alpha=0,...,3$.
This observation gives rise to the bilinear $\kappa$ RpV term.
Summarizing, in full RpV one has to take into account not only the usual
trilinear vertices but also the bilinear ones, which represent the
mixing between SUSY superfields. In the case of one loop diagrams the
bilinear insertions may be placed on the external neutrino lines
(neutrino--neutralino mixing) as well as inside the loop on the internal
fermion or boson line, or both. In this communication we concentrate on
the bilinear couplings inside the loop only, leaving the full discussion
to a~forthcoming regular paper.

\section{General diagrams leading to Majorana neutrino masses}

Except the usually discussed lepton--slepton and quark--squark
loops,\cite{mgozdz:nu_mm} one may construct new class of diagrams which
contain bilinear RpV insertions on the internal fermion or boson lines.
These diagrams give raise to the Majorana neutrino mass term exactly in
the same way as the usual ones.

The neutrino mass term will be proportional to the product of all the
coupling constants (two trilinear and two bilinear ones, the latter
having proper mass dimension) times the function coming from integration
over 4-momentum of all the propagators. To present the results in
a~compact form we first define two functions, ${\cal F}_1$ and ${\cal
  F}_2$ as follows:
\begin{eqnarray}
  {\cal F}_1(\alpha,\beta,\gamma,\delta) &=&
  \int_{-\infty}^{\infty} \frac{d^4k}{(2\pi)^4} 
  \frac{1}{(k^2-m_\alpha^2) (k^2 - m_\beta^2) 
    (k^2-m_\gamma^2) (k^2 - m_\delta^2)} \nonumber \\
  &=& \frac{1}{16\pi^2} \Bigg [
  \frac{m_\alpha^2} {(m_\alpha^2-m_\beta^2) (m_\alpha^2-m_\gamma^2)
    (m_\alpha^2-m_\delta^2)}
  \ln \left ( \frac{m_\alpha^2}{m_\gamma^2} \right ) \nonumber \\
  &+&
  \frac{m_\beta^2} {(m_\beta^2-m_\alpha^2) (m_\beta^2-m_\gamma^2)
    (m_\beta^2-m_\delta^2)}
  \ln \left ( \frac{m_\beta^2}{m_\gamma^2} \right ) \nonumber \\
  &+&
  \frac{m_\delta^2} {(m_\delta^2-m_\alpha^2) (m_\delta^2-m_\beta^2)
    (m_\delta^2-m_\gamma^2)} 
  \ln \left ( \frac{m_\delta^2}{m_\gamma^2} \right ) \Bigg ]
\end{eqnarray}
\begin{eqnarray}
  {\cal F}_2(\alpha,\beta,\gamma,\delta) &=&
  \int_{-\infty}^{\infty} \frac{d^4k}{(2\pi)^4} 
  \frac{k^2}{(k^2-m_\alpha^2) (k^2 - m_\beta^2) 
    (k^2-m_\gamma^2) (k^2 - m_\delta^2)} \nonumber \\
  &=& \frac{1}{16\pi^2} \Bigg [
  \frac{m_\alpha^4} {(m_\alpha^2-m_\beta^2) (m_\alpha^2-m_\gamma^2)
    (m_\alpha^2-m_\delta^2)}
  \ln \left ( \frac{m_\alpha^2}{m_\gamma^2} \right ) \nonumber \\
  &+&
  \frac{m_\beta^4} {(m_\beta^2-m_\alpha^2) (m_\beta^2-m_\gamma^2)
    (m_\beta^2-m_\delta^2)}
  \ln \left ( \frac{m_\beta^2}{m_\gamma^2} \right ) \nonumber \\
  &+&
  \frac{m_\delta^4} {(m_\delta^2-m_\alpha^2) (m_\delta^2-m_\beta^2)
    (m_\delta^2-m_\gamma^2)} 
  \ln \left ( \frac{m_\delta^2}{m_\gamma^2} \right ) \Bigg ]
\end{eqnarray}
The indices $\alpha, \beta, \dots$ label particles with masses
$m_\alpha, m_\beta, \dots$, respectively, appearing inside the loop. One
can check that both of theses functions are symmetric with respect to
permutation of any two masses.

We present the three new diagrams together with the corresponding loop
functions in Tab.~\ref{mgozdz:tab1} below. We use the convention that
fermions inside the loop are successively labeled by $f1$, $f2$, and
$f3$, and similarly the bosons by $b1$, $b2$, and $b3$. If the loop
contains only one particle of given spin, it is labeled simply by $f$ or
$b$.

\begin{table}[!h]
\tbl{\label{mgozdz:tab1}Loop integrals governing Majorana neutrino
  mass.}{
\begin{tabular}{ccc}
\toprule
& Diagram & Integral ${\cal I}$\\
\colrule
I &
\parbox{0.3\textwidth}{
\vskip 4 truept
\includegraphics[width=0.25\textwidth]{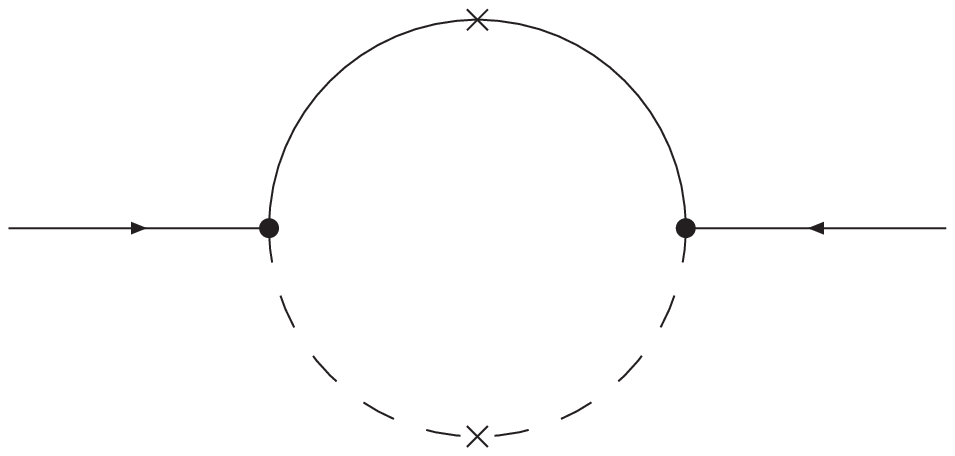}
\vskip 4 truept} &
\parbox{0.5\textwidth}{
\begin{eqnarray*}
m_{f1} m_{f2}\ {\cal F}_1(f1,f2,b1,b2) + {\cal F}_2(f1,f2,b1,b2)
\end{eqnarray*}} \\
II &
\parbox{0.3\textwidth}{
\vskip 4 truept
\includegraphics[width=0.25\textwidth]{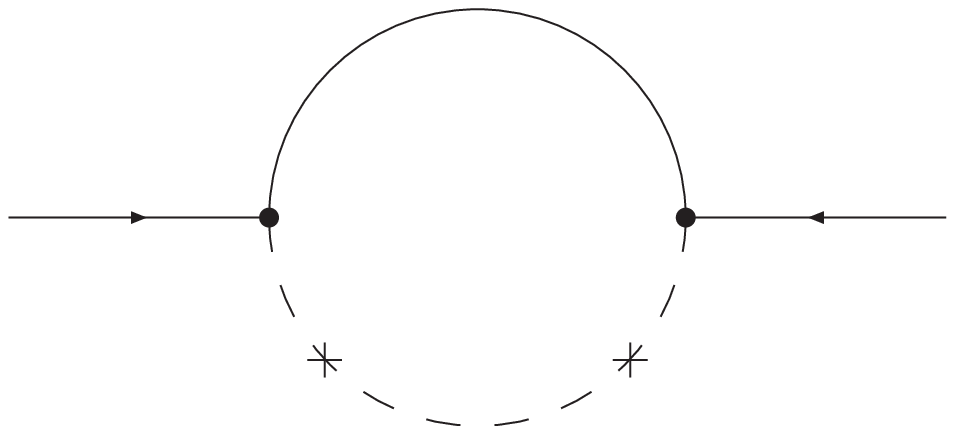}
\vskip 4 truept} &
\parbox{0.5\textwidth}{
\begin{eqnarray*}
m_f\ {\cal F}_1(f,b1,b2,b3)
\end{eqnarray*}} \\
III &
\parbox{0.3\textwidth}{
\vskip 4 truept
\includegraphics[width=0.25\textwidth]{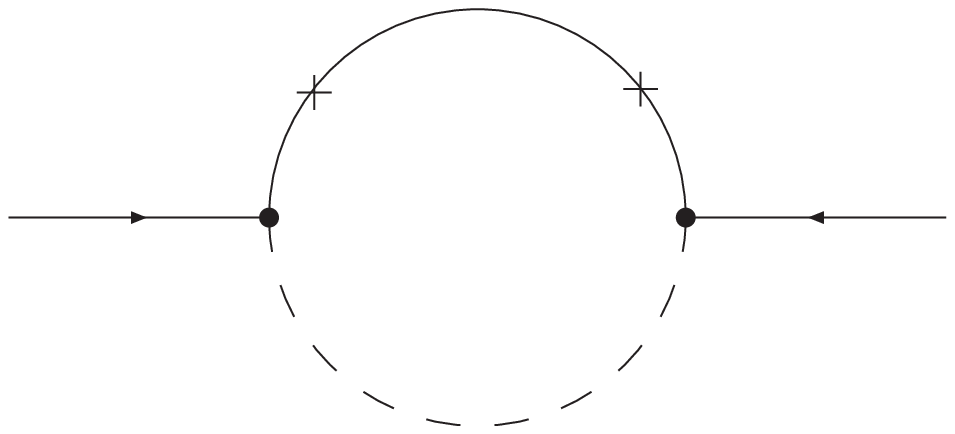}
\vskip 4 truept} &
\parbox{0.5\textwidth}{
\begin{eqnarray*}
&& m_{f1} m_{f2} m_{f3}\ {\cal F}_1(f1,f2,f3,b) \\
&+& (m_{f1} + m_{f2} + m_{f3})\ {\cal F}_2(f1,f2,f3,b)
\end{eqnarray*}} \\
\botrule
\end{tabular}}
\end{table}

\section{General diagrams leading to Majorana neutrino transition
  magnetic moments}

The magnetic moment of the Majorana neutrino is proportional to the
amplitude of an effective neutrino--photon--neutrino interaction.
Therefore an external photon has to be added to the diagrams considered
in the previous section. The photon may interact only with particles
possessing non-zero electric charge, so it may be attached only to the
internal lines of the loop. The external lines contain neutrinos or,
more exactly, a~neutrino-neutralino superposition, which is electrically
neutral.

As before, we start by calculating two integrals in the momentum space,
which represent the situation when the photon interacts with the
particle with mass $m_\alpha$:
\begin{eqnarray}
  {\cal F}_3(\alpha,\beta,\gamma,\delta) &=&
  \int_{-\infty}^{\infty} \frac{d^4k}{(2\pi)^4}
  \frac{1}{(k^2-m_\alpha^2)^2 (k^2 - m_\beta^2) 
    (k^2-m_\gamma^2) (k^2 - m_\delta^2)} \nonumber \\
  &=& -\frac{1}{2m_\alpha} \frac{\partial}{\partial m_\alpha}
  {\cal F}_1 (\alpha,\beta,\gamma,\delta),
\end{eqnarray}
\begin{eqnarray}
  {\cal F}_4(\alpha,\beta,\gamma,\delta) &=&
  \int_{-\infty}^{\infty} \frac{d^4k}{(2\pi)^4}
  \frac{k^2}{(k^2-m_\alpha^2)^2 (k^2 - m_\beta^2) 
    (k^2-m_\gamma^2) (k^2 - m_\delta^2)} \nonumber \\
  &=& -\frac{1}{2m_\alpha}  \frac{\partial}{\partial m_\alpha}
  {\cal F}_2 (\alpha,\beta,\gamma,\delta).
\end{eqnarray}

The resulting loop integrals are presented in Tab.~\ref{mgozdz:tab2}.
For the sake of simplicity, we have not taken into account the possible
permutations of particles inside the loop. For example, in the case
labeled ``IV'' the photon could be attached also to the fermion $f2$,
which means that one should add a~$+(f1 \leftrightarrow f2)$ term to the
expression listed in the table. An interesting observation is, that some
of the processes (``V'' and ``VII'') are strictly forbidden. The zeros
come out as an exact result of $k$ integration of an odd $k$-function.
\begin{table}[!h]
\tbl{\label{mgozdz:tab2}Loop integrals governing Majorana neutrino
  transition magnetic moments.}{
\begin{tabular}{ccc}
\toprule
& Diagram & Integral ${\cal I}$\\
\colrule
IV &
\parbox{0.3\textwidth}{
\vskip 4 truept
\includegraphics[width=0.25\textwidth]{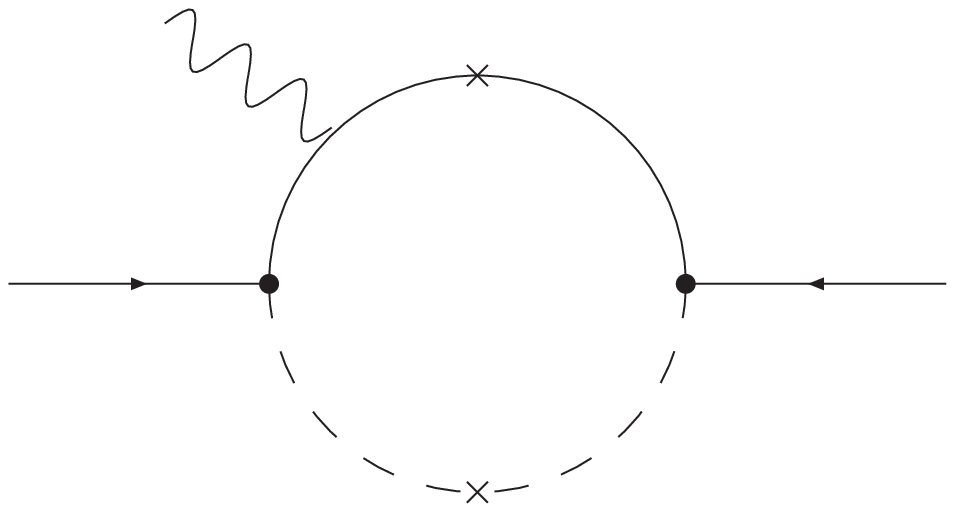}
\vskip 4 truept} &
\parbox{0.5\textwidth}{
\begin{eqnarray*}
&& m_{f1} m_{f2}\ {\cal F}_3(f1,f2,b1,b2) \\
&+& {\cal F}_4(f1,f2,b1,b2) 
\end{eqnarray*}} \\
V &
\parbox{0.3\textwidth}{
\vskip 4 truept
\includegraphics[width=0.25\textwidth]{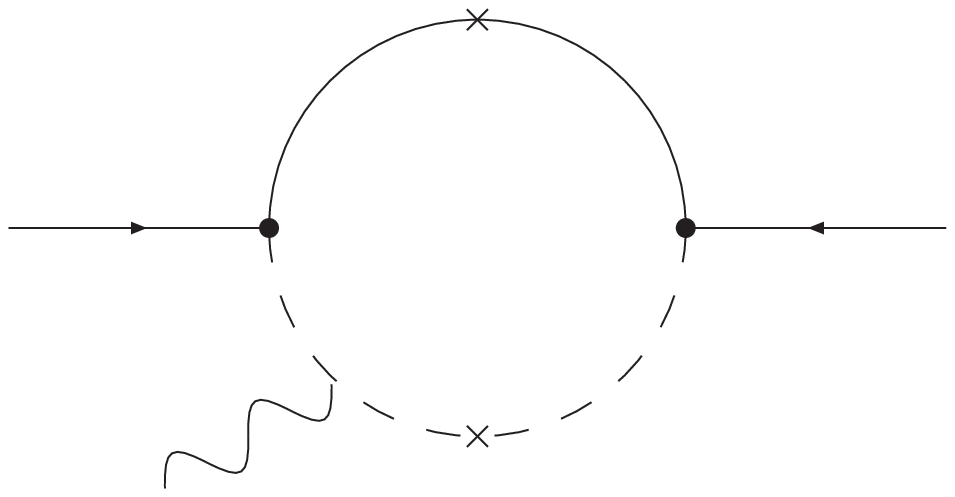}
\vskip 4 truept} &
\parbox{0.5\textwidth}{
\begin{eqnarray*}
0
\end{eqnarray*}} \\
VI &
\parbox{0.3\textwidth}{
\vskip 4 truept
\includegraphics[width=0.25\textwidth]{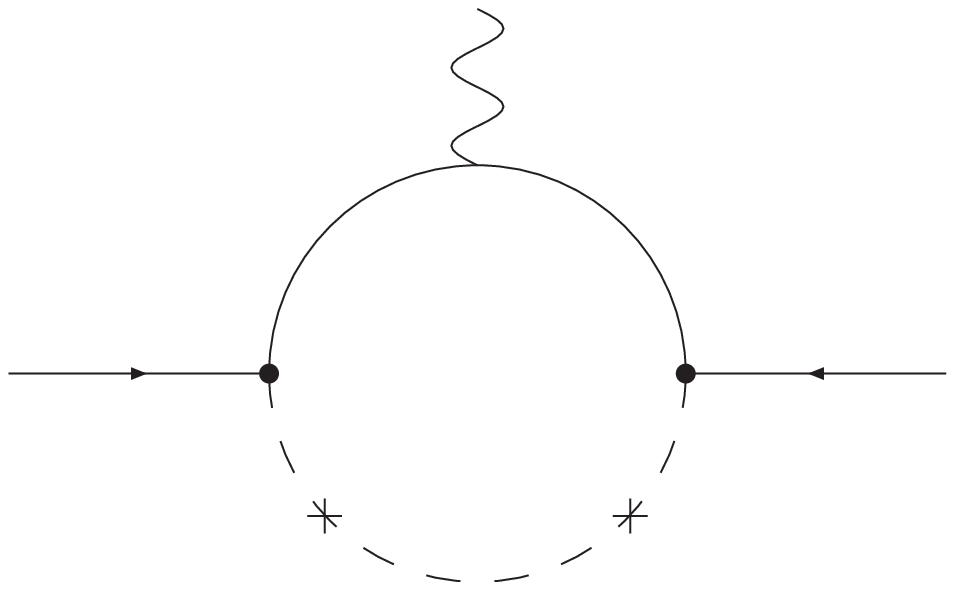}
\vskip 4 truept} &
\parbox{0.5\textwidth}{
\begin{eqnarray*}
m_f\ {\cal F}_4(f,b1,b2,b3)
\end{eqnarray*}} \\
VII &
\parbox{0.3\textwidth}{
\vskip 4 truept
\includegraphics[width=0.25\textwidth]{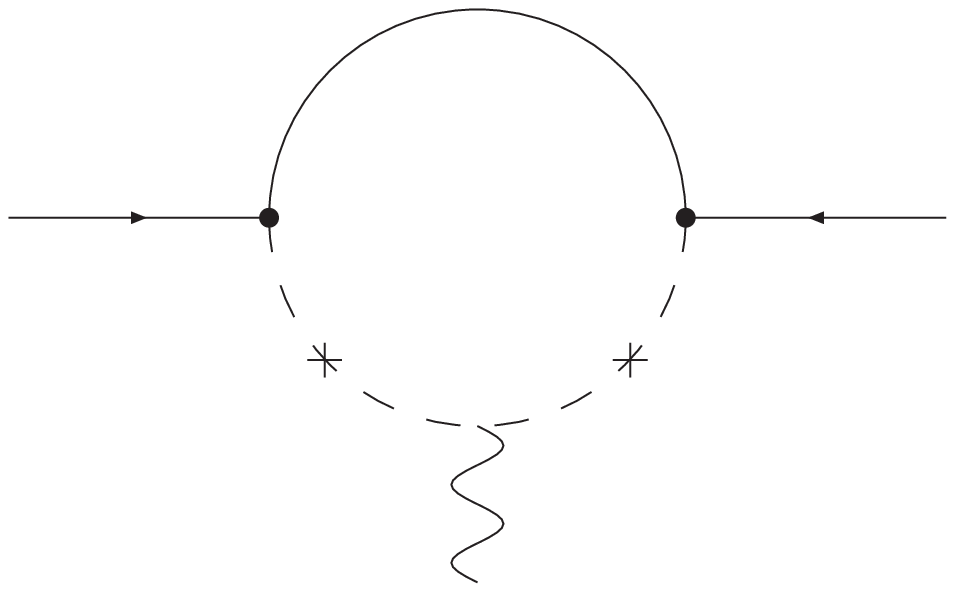}
\vskip 4 truept} &
\parbox{0.5\textwidth}{
\begin{eqnarray*}
0
\end{eqnarray*}} \\
VIII &
\parbox{0.3\textwidth}{
\vskip 4 truept
\includegraphics[width=0.25\textwidth]{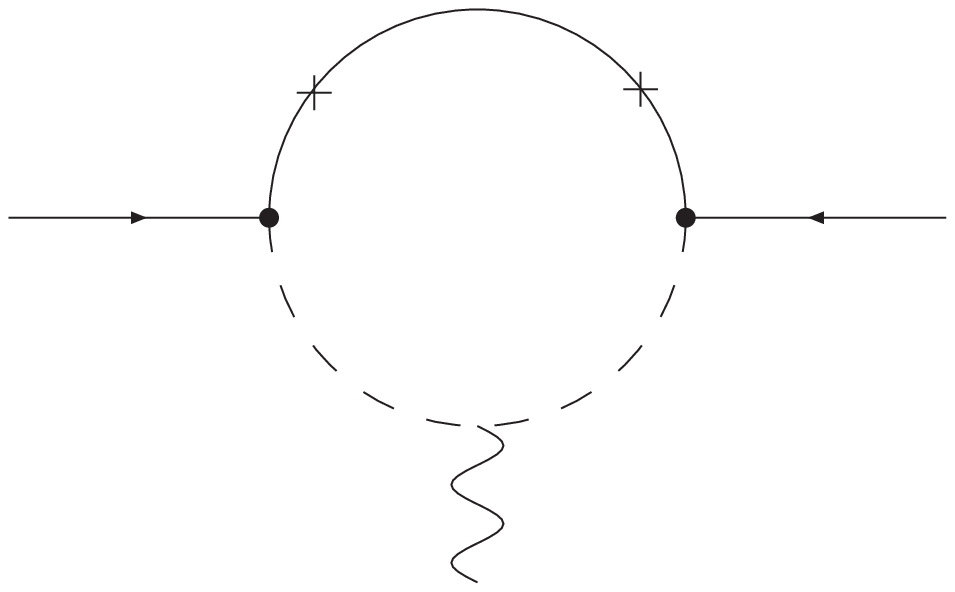}
\vskip 4 truept} &
\parbox{0.5\textwidth}{
\begin{eqnarray*}
4\ {\cal F}_4(b,f1,f2,f3)
\end{eqnarray*}} \\
IX &
\parbox{0.3\textwidth}{
\vskip 4 truept
\includegraphics[width=0.25\textwidth]{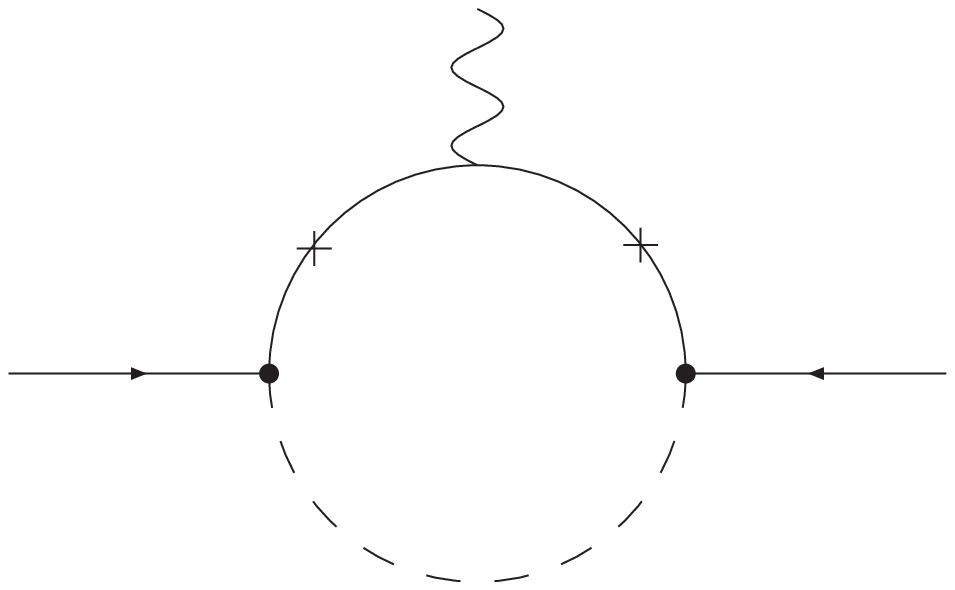}
\vskip 4 truept} &
\parbox{0.5\textwidth}{
\begin{eqnarray*}
&& m_{f1} m_{f2} (m_{f2}+m_{f3})\ {\cal F}_3(f2,f1,f3,b) \\
&+& (2m_{f1} + 3m_{f2} + m_{f3})\ {\cal F}_4(f2,f1,f3,b)
\end{eqnarray*}} \\
\botrule
\end{tabular}}
\end{table}

\section{Summary}
In this paper we have discussed the up-to-date problem of generating
Majorana neutrino mass and transition magnetic moments. The well known
mechanism, which exists within supersymmetric standard model with
non-conserved $R$-parity, has been extended to include also processes
described by bilinear terms in the superpotential. We have limited
ourselves to the case when bilinear insertions are present on the
internal lines of the fermion--boson loop. The remaining possibility,
where the insertions are on the external neutrino lines, are possible
due to the fact that in RpV models neutrinos mix with neutralinos. We
will discuss this problem in details elsewhere.

The lowest level diagrams of interest have been solved and the analytic
formulae can be found in Tab.~\ref{mgozdz:tab1} and~\ref{mgozdz:tab2}.
Using these functions the resulting mass $\cal M$ and magnetic moment
$\mu$ can be written as follows:
\begin{eqnarray}
  {\cal M} &=& c\ C_1 C_2 X_1 X_2\ {\cal I}, \\
  \mu &=& c\ C_1 C_2 X_1 X_2\ Q\ {\cal I}\ 2m_e\ \mu_B,
\end{eqnarray}
where the dimensionless trilinear coupling constants have been denoted
by $C_{1,2}$, the bilinear coupling constants by $X_{1,2}$, $\cal I$ is
the appropriate integral function, $\mu_B$ is the Bohr magneton, and
$m_e$ the electron mass. $c$ is the color factor and is equal to 3
whenever quarks or squarks appear in the loop; it is equal to 1 in all
other cases. $Q$ is a~dimensionless number and denotes the electric
charge of the particle to which the photon is attached, in units of the
elementary charge $e$.

It is not difficult to check that bilinear couplings appearing on
fermion lines have exactly the dimension of mass ($m$) and those on
boson lines have dimension mass squared ($m^2$).

\section*{Acknowledgments} 
The first author (MG) is partially supported by the Polish State
Committee for Scientific Research, and by the Foundation for Polish
Science. He would like also to express his gratitude to
prof.~A.~F\"a{\ss}ler for his warm hospitality in T\"ubingen during the
Summer 2006, and to prof. F. \v Simkovic for help in correcting the
manuscript.



\end{document}